\documentclass[aps,prl,reprint,superscriptaddress,longbibliography,nofootinbib,floatfix]{revtex4-2}

\usepackage{amsmath,amssymb,bm}
\usepackage{graphicx}
\usepackage{microtype}
\usepackage{xcolor}
\usepackage[colorlinks=true,allcolors=blue!45!black]{hyperref}
\usepackage{comment}

\makeatletter
\renewcommand{\frontmatter@abstractfont}{%
  \normalfont\normalsize
  \parindent=1em\relax
  \adjust@abstractwidth
}
\makeatother

\graphicspath{{figures/}}

\newcommand{\Tr}{\operatorname{Tr}}
\newcommand{\Diss}[1]{\mathcal{D}[#1]}
\newcommand{\cL}{\mathcal{L}}
\newcommand{\cE}{\mathcal{E}}
\newcommand{\ketbra}[2]{|#1\rangle\!\langle#2|}
\newcommand{\proj}[1]{|#1\rangle\!\langle#1|}

\begin{document}

\title{Entanglement Mpemba Effect}

\author{Ruicheng Bao}
\email{ruicheng@g.ecc.u-tokyo.ac.jp}
\affiliation{Department of Physics, Graduate School of Science, The University
of Tokyo, Hongo, Bunkyo-ku, Tokyo 113-0033, Japan}

\author{Yue Liu}
\email{yue.liu@yukawa.kyoto-u.ac.jp}
\affiliation{Center for Gravitational Physics and Quantum Information, Yukawa
Institute for Theoretical Physics, Kyoto University, Kitashirakawa Oiwakecho,
Sakyo-ku, Kyoto 606-8502, Japan}

\begin{abstract}
Generating entanglement rapidly and reliably is essential for quantum information processing, communication, and metrology. Dissipative preparation is attractive because engineered reservoirs robustly drive a system toward an entangled target, yet relaxation can carry a substantial time cost. Here we formulate the entanglement Mpemba effect, whereby an initially less entangled state overtakes a more entangled state under the same open-system dynamics. This effect turns initial-state engineering into a route for faster preparation without altering the dissipative protocol. We derive a general criterion for the reversal from the relaxation spectrum, applicable even when entanglement evolves nonmonotonically. A reversal of deterministic local operations and classical communication (LOCC)-reachability preorder provides a measure-independent certificate of reversed entanglement order. Exactly solvable %Bell- and GHZ-state preparation 
models show that initial-state selection can substantially shorten the time required to reach high entanglement. We further propose an experimentally relevant trapped-ion protocol that can realize the entanglement Mpemba effect.%keeps all dissipative operations fixed and changes only the initial state.
\end{abstract}

\maketitle

\textit{Introduction---}Entanglement is a central resource for quantum technologies.  It enables
quantum teleportation, entanglement-based cryptography, distributed quantum
networks, measurement-based computation, and quantum-enhanced metrology
\cite{Bennett1993,Ekert1991,Kimble2008,Raussendorf2001,
Giovannetti2011,Horodecki2009}.  Preparing highly entangled states with high
fidelity remains difficult because entangling operations must compete with
decoherence and imperfect control.  Reservoir engineering offers a robust
route in which the desired state is an attractor of an autonomous open-system
evolution \cite{Kraus2008,Verstraete2009,Barreiro2011,Mueller2011,
Lin2013,Schindler2013}.  The preparation nevertheless has a time cost.
Entanglement dynamics obey fundamental speed limits \cite{Pandey2024}, and the
relaxation time of dissipative protocols can grow as the target becomes more
entangled \cite{Pocklington2024}.  Optimized coherent control and adaptive
reservoirs can shorten this time \cite{Sorelli2019,Pocklington2025}, although
they require additional operations or changes to the dynamics during
preparation.  Finding faster routes under a fixed relaxation law is therefore
an important and less explored problem.

An elegant alternative is initial-state engineering.  One keeps the reservoir and external protocol fixed, chooses an initial state
with favorable kinetics, and lets the same robust relaxation prepare the
target.  Initial-state-dependent speed limits already show that this choice
can change the time required for dissipative state preparation
\cite{LiuNie2023}.  Once the reservoir has been engineered, no time-dependent
or adaptive modification is needed during relaxation.  The Mpemba effect
provides a natural framework for this idea.  In its original setting, hotter
water can cool or freeze before colder water under the same conditions
\cite{mpemba1969cool}.  More generally, a state that initially appears farther
from the final state can overtake a closer state and arrive first.  For
Markovian dynamics, the initial state fixes the amplitudes of the decaying
generator modes \cite{lu2017nonequilibrium,klich2019mpemba}. If the slowest
relaxation mode is absent for an initial state, relaxation is governed by a faster rate and
the late-time errors become exponentially small compared to other initial states. This is the strong
Mpemba effect \cite{lu2017nonequilibrium,klich2019mpemba}, and such exponentially accelerated cooling
has been observed in a colloidal system \cite{kumar2020exponentially}.
Relaxation in metastable landscapes \cite{Kumar22,walker2021anomalous,chetrite2021metastable} and confined systems \cite{liu2026mpemba,pre-us} further show
that the apparent initial distance does not determine the relaxation time.  Related behavior
has been found in molecular and granular gases, inertial suspensions, driven
matter, nonequilibrium steady states, and active systems
\cite{Santos17,Torrente19,Biswas20,Santos20,Takada21a,
Gonzalez2021,Degunther2022,Biswas24}.  %The anomaly is fundamentally important because it reverses the expected ordering of relaxation, and it is practically useful because the acceleration can come entirely from the initial state.

The quantum counterpart of this phenomenon has been studied in Markovian open systems,
reservoir-coupled few-level systems, exceptional-point dynamics,
nonequilibrium reservoirs, non-Markovian evolution, and many-body symmetry
restoration \cite{Carollo21,Ivander23,Chatterjee_2023,Chatterjee2024,
Moroder2024,Wang24,Nava24,Strachan2025,Turkeshi2025,Bao2025, Bao2026}. Ordinary and strong
quantum Mpemba effects have also been observed experimentally
\cite{Joshi2024,Shapira2024,Zhang2025}, and recent reviews summarize this
rapidly developing field \cite{AresReview2025,Yu2025,TEZA2026}. These developments raise a direct question about dissipative entanglement preparation. 
\textit{Can a less entangled
initial state overtake a more entangled one while both evolve toward the same
entangled target, and can this reversal reduce the time needed to reach a
useful entanglement threshold?} An affirmative answer would provide a
practical shortcut based only on initial-state selection.

We give this affirmative answer and introduce the entanglement Mpemba effect.
For two initial states evolving under the same open-system dynamics, the
effect occurs when the initially less entangled state later becomes the
more entangled one.  We derive a sufficient criterion for the reversal from the relaxation spectrum, applicable even when entanglement evolves nonmonotonically. We use deterministic LOCC reachability to certify a reversed ordering simultaneously for all entanglement monotones, with equality allowed for some monotones.
We distinguish
this ordering reversal from a preparation advantage through first-hitting
times. Exactly solvable Bell- and Greenberger–Horne–Zeilinger (GHZ)-state preparation models show how a less entangled initial state accesses a faster relaxation sector, exhibits the reversal under both concurrence-based and negativity-based measures, and reaches a high-entanglement threshold earlier. Finally, we propose an experimentally relevant
trapped-ion protocol in which only the initial state is changed. These results reveal a fundamental anomaly in entanglement relaxation and establish initial-state selection as a direct route to faster dissipative entanglement preparation.

\textit{Setup---}Fix a finite-dimensional multipartite system and a partition.  Let
\(\rho_j(t)\) be density operators at time
\(t\geq0\), with \(j\in\{A,B\}\) labeling the two preparations, and let both
evolve under the same time-homogeneous completely positive trace-preserving
semigroup
\begin{equation}
 \rho_j(t)=\Phi_t[\rho_j(0)]=e^{t\cL}\rho_j(0),
 \qquad j=A,B ,
 \label{eq:dynamics}
\end{equation}
where \(\Phi_t\) is the dynamical map and \(\cL\) its
Gorini--Kossakowski--Sudarshan--Lindblad generator
\cite{Lindblad1976,Gorini1976}.  We assume
\(\rho_j(t)\to\rho_\star\), with \(\cL\rho_\star=0\), in their common
accessible sector.  Let \(E\) be a continuous entanglement monotone for the
chosen partition and define
\(E_j(t)\equiv E[\rho_j(t)]\) and
\(E_\star\equiv E(\rho_\star)\).  The ordering difference is
\begin{equation}
 \Delta_E(t)=E_A(t)-E_B(t).
 \label{eq:delta}
\end{equation}
An \(E\)-entanglement Mpemba effect occurs when
\(\Delta_E(0)<0\) but \(\Delta_E(t)>0\) on a later interval.  For continuous
\(E_j(t)\), a crossing event is an isolated zero or a maximal zero interval
that separates regions of opposite sign.  A zero with the same sign on both
sides is a contact.  The effect therefore requires a crossing of the two
entanglement trajectories.  For an entanglement threshold \(e\), define
the first hitting time
\begin{equation}
 \tau_E(\rho;e)=\inf\{t\geq0:E[\Phi_t(\rho)]\geq e\},
 \label{eq:hitting}
\end{equation}
with \(\tau_E=+\infty\) if the threshold is never reached.  Ordering reversal
and first-hitting advantage are distinct.  Near a transverse upward crossing
{at which \(E_A\) is locally increasing,}
\(A\) reaches {sufficiently nearby thresholds above the crossing level}
first if neither trajectory visited them earlier.  Neither definition refers to fidelity or assumes a sign for
\(\dot E_j(t)=dE_j(t)/dt\).

\textit{Entanglement-visible spectrum---}First suppose that \(\cL\) is diagonalizable on the accessible trace-zero
operator space.  Let \(R_\alpha\) and \(L_\alpha\) denote right and left
eigenoperators, normalized by
\(\Tr(L_\alpha^\dagger R_\beta)=\delta_{\alpha\beta}\), with
\(\cL R_\alpha=\lambda_\alpha R_\alpha\) and
\(\cL^\dagger L_\alpha=\lambda_\alpha^*L_\alpha\).  Excluding the stationary mode,
the trajectories have the Liouville expansion
\begin{equation}
 \rho_j(t)=\rho_\star+\sum_{\alpha\geq1}
 c_{j\alpha}e^{\lambda_\alpha t}R_\alpha ,
 \qquad \Re\lambda_\alpha<0 .
 \label{eq:liouville}
\end{equation}
The coefficient
\(c_{j\alpha}=\Tr\{L_\alpha^\dagger[\rho_j(0)-\rho_\star]\}\) gives the
initial-state amplitude of mode \(\alpha\).  Suppose that both trajectories
eventually enter a common sector in which, as \(X\to0\) along accessible
trace-zero directions, \(E\) admits the asymptotic directional expansion
\begin{equation}
 E(\rho_\star+X)\sim E_\star+
 \sum_{m\geq1}Q_m(X,\ldots,X).
 \label{eq:directional}
\end{equation}
The map \(Q_m\) is the real symmetric \(m\)-linear directional response on
Hermitian perturbations.  We use the same symbol for its unique
complex-multilinear extension.  Conjugate tuples are grouped so that
\(\Delta_E\) is real.  For the
multi-index
\(\bm\alpha=(\alpha_1,\ldots,\alpha_m)\), introduce
\(q_{\bm\alpha}^{(m)}
=Q_m(R_{\alpha_1},\ldots,R_{\alpha_m})\),
\(c_{j,\bm\alpha}^{(m)}=\prod_{k=1}^{m}c_{j\alpha_k}\), and
\(\Lambda_{\bm\alpha}^{(m)}
=\sum_{k=1}^{m}\lambda_{\alpha_k}\).  Substitution of
Eq.~\eqref{eq:liouville} gives
\begin{equation}
 \Delta_E(t)\sim
 \sum_{m\geq1}\sum_{\bm\alpha}
 q_{\bm\alpha}^{(m)}
 \left[c_{A,\bm\alpha}^{(m)}-c_{B,\bm\alpha}^{(m)}\right]
 e^{\Lambda_{\bm\alpha}^{(m)}t}.
 \label{eq:visible-expansion}
\end{equation}
The symbol \(\sim\) denotes an asymptotic series as \(t\to\infty\).
Terms with equal exponent sums must be grouped before their coefficients are
tested.  The nonzero grouped terms form the entanglement-visible comparison
spectrum.  Both the initial amplitudes and the responses \(Q_m\) determine
which terms survive.  In particular, a slow Liouvillian mode may carry zero
initial amplitude or lie tangent to an entanglement level set.  It may also
enter first through a product whose total decay rate is faster.
A linear target overlap (equivalently, fidelity for a pure
target) retains only the \(m=1\) response to one fixed functional.

For a piecewise-smooth monotone, Eq.~\eqref{eq:visible-expansion} applies on a
common branch that admits this expansion.  Other nonsmooth approaches require
trajectory-specific asymptotics, while Jordan blocks add powers of \(t\)
(see the Supplemental Material (SM){~\cite{SupplementalMaterial})}.
The dominant visible rate and amplitude
can be inferred from the late-time comparison signal without full Liouvillian
tomography.  A fit to the target population alone may miss a visible product
of modes. A nonzero leading coefficient is typically stable under small
perturbations.  %\textcolor{red}{On any fixed finite observation window, a positivelead with a nonzero margin persists under sufficiently small generator perturbations, as quantified in the SM.}

\textit{Spectral criterion for the entanglement Mpemba effect---}For simplicity, we first consider the case in which, after grouping degenerate terms, the dominant contribution is nonoscillatory and has the form
\begin{equation}
 \Delta_E(t)=K_Et^{q_E} e^{-r_Et}[1+o(1)]\textcolor{red}{,}
 \label{eq:criterion}
\end{equation}
Here \(K_E\in\mathbb R\setminus\{0\}\) is the real grouped amplitude,
\(r_E>0\) the leading entanglement-visible decay rate,
\(q_E\in\mathbb N_0=\{0,1,\ldots\}\) the surviving Jordan-block power, and
\(o(1)\to0\) as \(t\to\infty\).
Then
\begin{equation}
 \Delta_E(0)<0,\qquad K_E>0
 \label{eq:sign-condition}
\end{equation}
is sufficient for an entanglement Mpemba effect.  State \(A\) then remains more
entangled than \(B\) at all sufficiently late times.  Equation
\eqref{eq:criterion} fixes the late-time sign, and continuity guarantees at
least one ordering-reversing crossing event between the initial and
asymptotic regimes.

The criterion is sufficient rather than necessary and does not imply a
unique crossing.  If the zero set has finitely many components, opposite
endpoint signs imply an odd number of sign-changing events.
More generally, the leading layer has the form
\(t^{q_E}e^{-r_Et}[F_E(t)+o(1)]\), where \(F_E(t)\) is a real trigonometric
polynomial, see SM for details.

{
\textit{Measure-independent LOCC order---}The sign criterion above resolves
one specified monotone \(E\).  With respect to the fixed partition, define the
deterministic-LOCC reachability preorder for arbitrary pure or mixed,
bipartite or multipartite states by
\begin{equation}
 \rho\succeq_{\rm LOCC}\sigma
 \quad\Longleftrightarrow\quad
 \exists\,\Lambda\in{\rm LOCC}:\ \Lambda(\rho)=\sigma .
 \label{eq:LOCC-order}
\end{equation}
Every entanglement monotone is nonincreasing under deterministic LOCC, and
hence
\begin{equation}
 \rho\succeq_{\rm LOCC}\sigma
 \quad\Longrightarrow\quad
 M(\rho)\geq M(\sigma)\quad\forall\,M .
 \label{eq:all-monotone-order}
\end{equation}
Thus, if \(\rho_B(0)\succeq_{\rm LOCC}\rho_A(0)\) while
\(\rho_A(t)\succeq_{\rm LOCC}\rho_B(t)\) later, every monotone has the
reversed weak ordering.  Strict crossing for a chosen monotone still requires
strict scalar inequalities.  This certificate is sufficient rather than
necessary for reversal of a chosen monotone and remains valid for mixed and multipartite states
\cite{Horodecki2009}.  For bipartite pure states it reduces, by Nielsen's
theorem, to Schmidt majorization \cite{Nielsen1999,Vidal2000}. The argument here is similar in spirit to the thermomajorization Mpemba effect \cite{Tan2025}. %A mixed-state realization is given after the two-qubit example and derived in the SM.
\par}

\textit{Preparation-time advantage---}When both trajectories eventually approach \(E_\star\) from below, define
the positive deficits \(D_j(t)\equiv E_\star-E_j(t)\).  Suppose their leading
late-time terms are simple nonoscillatory exponentials,
\begin{equation}
 D_j(t)=A_je^{-r_jt}[1+o(1)],
 \qquad A_j>0.
 \label{eq:deficits}
\end{equation}
Here \(A_j\) and \(r_j>0\) are the leading deficit amplitude and decay rate
for preparation \(j\).
Since \(\Delta_E=D_B-D_A\), the late ordering is reversed when
\(r_A>r_B\), or, for equal rates, when \(A_A<A_B\).  If sufficiently high
thresholds were not reached earlier, then, with
\(\tau_j(e)\equiv\tau_E[\rho_j(0);e]\),
\(\tau_j(e)=r_j^{-1}\ln[A_j/(E_\star-e)]+o(1)\) as
\(e\uparrow E_\star\).  Under this no-earlier-visit condition, a larger visible
rate produces a growing first-hitting-time advantage as the target approaches
\(E_\star\).

\begin{figure}[t]
 \centering
 \includegraphics[width=0.95\columnwidth]{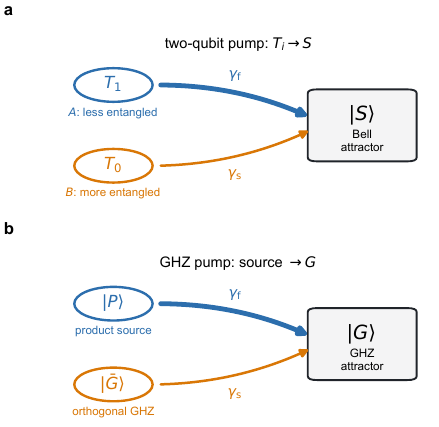}
 \caption{(a) A two-qubit generator pumps the product source \(T_1\) rapidly and the
 orthogonal Bell source \(T_0\) slowly into the singlet.
 (b) The same construction pumps a product source and an orthogonal GHZ
 state into a common GHZ attractor.  In both cases all channels are present
 for both preparations; the initial support selects the active one.}
 \label{fig:schematics}
\end{figure}

\paragraph*{\textit{Two kinetic realizations.---}}
Figure~\ref{fig:schematics} shows the construction shared by the two
examples.  A fixed generator contains fast and slow source-to-target pumps.
Orthogonal initial supports select the active face: the less-entangled state
\(A\) occupies the fast face, whereas the more-entangled state \(B\) occupies
the slow one.
For the two-qubit realization define
\(|S\rangle=(|01\rangle-|10\rangle)/\sqrt2\),
\(|T_0\rangle=(|01\rangle+|10\rangle)/\sqrt2\), and
\(|T_1\rangle=|00\rangle\).  For the three-qubit realization define
\(|G\rangle=(|000\rangle-|111\rangle)/\sqrt2\),
\(|\bar G\rangle=(|000\rangle+|111\rangle)/\sqrt2\), and
\(|P\rangle=|001\rangle\).  The subscripts \({\rm f}\) and \({\rm s}\)
denote the fast and slow population-decay rates
\(\gamma_{\rm f}>\gamma_{\rm s}>0\).

\textit{Two-qubit result---}For these states, the fixed pump is
\begin{equation}
 \cL\rho=
 \Diss{\sqrt{\gamma_{\rm s}}\ketbra{S}{T_0}}\rho+
 \Diss{\sqrt{\gamma_{\rm f}}\ketbra{S}{T_1}}\rho ,
 \qquad \gamma_{\rm f}>\gamma_{\rm s}.
 \label{eq:pump}
\end{equation}
Here
\(\Diss{L}\rho=L\rho L^\dagger-\tfrac12\{L^\dagger L,\rho\}\) is the
Lindblad dissipator.  For initial singlet weights \(a,b\in[0,1]\), the pump
acts on
\begin{align}
 \rho_A(0)&=a\proj{S}+(1-a)\proj{T_1},\nonumber\\
\rho_B(0)&=b\proj{S}+(1-b)\proj{T_0}.
 \label{eq:initial}
\end{align}

\begin{figure}[!t]
 \centering
 \includegraphics[width=0.95\columnwidth]{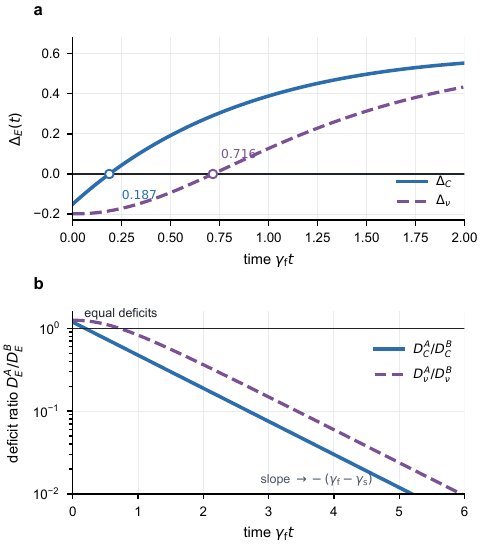}
 \caption{(a) Concurrence and normalized negativity reverse their initial ordering at
 different times under the same generator.
 (b) The ratios of the individual deficits decay with the visible-rate
 difference \(\gamma_{\rm f}-\gamma_{\rm s}\), as predicted by the spectral
 criterion.}
 \label{fig:two-qubit-results}
\end{figure}

We consider the branch \(1/2<b<1\) and \(0\leq a<2b-1\), for which
\(A\) initially has the lower concurrence.  Each source face is invariant.
The \(T_0\) dissipator annihilates the entire \(A\) trajectory, and the
\(T_1\) dissipator annihilates the entire \(B\) trajectory.  Writing the
singlet populations as
\(p_j(t)\equiv\langle S|\rho_j(t)|S\rangle\) gives
\begin{equation}
 p_A(t)=1-(1-a)e^{-\gamma_{\rm f}t},\qquad
 p_B(t)=1-(1-b)e^{-\gamma_{\rm s}t}.
 \label{eq:populations}
\end{equation}

Both jump operators remain present in every run.  An unoccupied source simply
has zero jump probability.  Weak mixing between the faces exposes both
exponentials but leaves the reversal intact over a finite perturbative range
(SM).

We use the two-qubit concurrence \(C(\rho)\) and the negativity
\(\mathcal N(\rho)=[\|\rho^{\Gamma_1}\|_1-1]/2\), where
\(\Gamma_1\) denotes partial transpose on the first qubit and
\(\|\cdot\|_1\) the trace norm
\cite{Wootters1998,VidalWerner2002}.  The normalized negativity is
\(\nu(\rho)\equiv2\mathcal N(\rho)\).  With
\(C_j(t)\equiv C[\rho_j(t)]\) and
\(\nu_j(t)\equiv\nu[\rho_j(t)]\), the two measures reduce to
\begin{align}
 C_A&=p_A,& C_B&=2p_B-1,\nonumber\\
 \nu_A&=\sqrt{(1-p_A)^2+p_A^2}-(1-p_A),&
 \nu_B&=2p_B-1 .
 \label{eq:monotones}
\end{align}
The two measures are inequivalent along the fast path:
\(C_A=p_A\), whereas
\(\nu_A=p_A^2/2+O(p_A^3)\) near \(p_A=0\).  Close to the singlet, however,
their deficits inherit the fast rate for \(A\) and the slow rate for \(B\).
The spectral criterion therefore predicts a reversal for both measures.

For
\(a=0.05\), \(b=0.60\), and
\(\gamma_{\rm s}/\gamma_{\rm f}=0.08\), we measure time in units of
\(\gamma_{\rm f}^{-1}\).  State \(A\) starts below \(B\) in concurrence
\((0.05<0.20)\) and normalized negativity \((0.0013<0.20)\).  Yet
\(\Delta_C(t)\equiv C_A(t)-C_B(t)\) and
\(\Delta_\nu(t)\equiv\nu_A(t)-\nu_B(t)\)
change sign at \(\gamma_{\rm f}t=0.187\) and \(0.716\), respectively
[Fig.~\ref{fig:two-qubit-results}(a)].  Because the crossing times differ, no
single linear target population accounts for both reversals.

The exact solution also gives the preparation-time gain.  For a concurrence
threshold \(\max\{a,2b-1\}<c<1\), define
\(\tau_C^j(c)\equiv\tau_C[\rho_j(0);c]\).  Then
\begin{equation}
 \tau_C^A(c)=\frac{1}{\gamma_{\rm f}}
 \ln\frac{1-a}{1-c},\qquad
 \tau_C^B(c)=\frac{1}{\gamma_{\rm s}}
 \ln\frac{2(1-b)}{1-c}.
 \label{eq:thresholds}
\end{equation}
At \(c=0.9\), the preparation times are \(2.251\) and \(25.993\), an
\(11.55\)-fold speedup.  At \(\nu=0.9\), the corresponding speedup is
\(8.87\).  Figure~\ref{fig:two-qubit-results}(b) displays the corresponding
asymptotic test.  For \(X\in\{C,\nu\}\), define
\(D_X^j(t)\equiv1-X_j(t)\).  Both ratios \(D_X^A/D_X^B\) fall below unity,
and their logarithmic slope approaches
\(-(\gamma_{\rm f}-\gamma_{\rm s})\), as predicted by
Eq.~\eqref{eq:deficits}.

\begin{comment}

{\color{red} The \(|T_0\rangle\) and \(|\bar G\rangle\) pumps plotted here illustrate monotone-dependent reversals.
A separate product-source variant of the Bell pump realizes the stronger
LOCC-order reversal while retaining the population laws in
Eq.~\eqref{eq:populations}.  Replace the slow source \(|T_0\rangle\) by
\(|11\rangle\) in both its jump and initial state.  Because \(X\otimes X\)
fixes \(\proj{S}\) and exchanges \(\proj{00}\) with \(\proj{11}\), while
deterministic local resets reduce the singlet weight, the LOCC direction
reverses exactly at
\begin{equation}
 t_{\rm X}=\frac{\ln[(1-a)/(1-b)]}
 {\gamma_{\rm f}-\gamma_{\rm s}} .
 \label{eq:LOCC-crossing-time}
\end{equation}
Explicit conversion channels and the multipartite analogue are given in the
SM. 
\par}
\end{comment}

\begin{figure}[t]
 \centering
 \includegraphics[width=0.95\columnwidth]{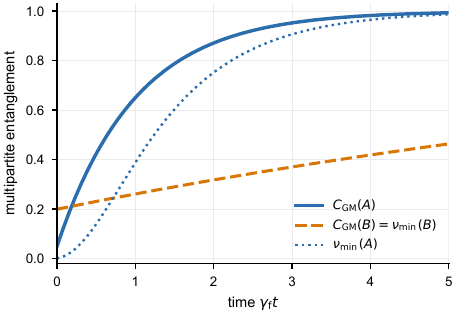}
 \caption{The initially less-entangled fast state overtakes the slow state for both
 genuine multipartite concurrence and the minimum normalized one-versus-two
 cut negativity.  The two measures follow inequivalent fast branches.}
 \label{fig:ghz-results}
\end{figure}

\textit{Multipartite result---}The same channel-selection mechanism applies beyond bipartite concurrence.
Replace the jumps
in Eq.~\eqref{eq:pump} with
\(\sqrt{\gamma_{\rm s}}\ketbra{G}{\bar G}\) and
\(\sqrt{\gamma_{\rm f}}\ketbra{G}{P}\).  The resulting generator has the two
invariant faces in Fig.~\ref{fig:schematics}(b).  Additional pumps from unused basis states can
make \(|G\rangle\) globally attractive without affecting either trajectory.

For
\(\rho_A(t)=p_A(t)\proj{G}+[1-p_A(t)]\proj{P}\) and
\(\rho_B(t)=p_B(t)\proj{G}+[1-p_B(t)]\proj{\bar G}\), with the populations
from Eq.~\eqref{eq:populations}, we use the genuine multipartite concurrence
\(C_{\rm GM}\) \cite{Rafsanjani2012} and the across-all-cuts diagnostic
\(\nu_{\min}(\rho)\equiv\min_{k=1,2,3}2\mathcal N_{k|\bar k}(\rho)\).
Here \(\mathcal N_{k|\bar k}\) is the negativity across the bipartition of
qubit \(k\) from the remaining two qubits.  Since \(p_B(t)\geq b>1/2\), the
exact branches are
\begin{align}
 C_{\rm GM}^A&=p_A, & C_{\rm GM}^B&=2p_B-1,
 \nonumber\\
 \nu_{\min}^A&=\sqrt{(1-p_A)^2+p_A^2}-(1-p_A),
 & \nu_{\min}^B&=2p_B-1 .
 \label{eq:ghz-cgm}
\end{align}
The two multipartite quantifiers inherit the same distinct crossing times as
the two-qubit measures [Fig.~\ref{fig:ghz-results}].  Their comparison spectra
depend on the chosen partition and on the resource geometry, rather than on a
special property of Wootters concurrence.

\textit{Fixed-cycle realization---}We propose a digital trapped-ion realization of the two-qubit pump using two
system ions and one resettable ancilla.  For \(i\in\{{\rm s},{\rm f}\}\), define
\(|T_{\rm s}\rangle\equiv|T_0\rangle\) and
\(|T_{\rm f}\rangle\equiv|T_1\rangle\).  A source-selective transfer with
probability \(p_i\), followed by ancilla reset, implements the Kraus operators
\begin{equation}
 K_0^{(i)}=\mathbb I+(\sqrt{1-p_i}-1)\proj{T_i},\qquad
 K_1^{(i)}=\sqrt{p_i}\ketbra{S}{T_i}.
 \label{eq:kraus}
\end{equation}
Here \(\mathbb I\) is the system identity, and these Kraus operators define a
completely positive trace-preserving block \(\cE_i\).  Every shot applies the
same cycle \(\cE=\cE_{\rm f}\circ\cE_{\rm s}\), including both blocks and
resets.  Only the initial system state changes.  After \(n\in\mathbb N_0\) cycles, each
source population is multiplied by \((1-p_i)^n\), exactly reproducing the
continuous pump under \(\gamma_i=-\delta t^{-1}\ln(1-p_i)\), where
\(\delta t\) is one cycle duration.  For \(p_{\rm f}=0.25\) and
\(p_{\rm s}=0.05\), the ideal map reaches
\(C=0.9\) in \(8\) rather than \(41\) cycles.  Trapped-ion experiments have
demonstrated digital open-system maps and dissipative Bell-state preparation
\cite{Barreiro2011,Schindler2013,Lin2013,MolmerSorensen1999}.  The details are
%\textcolor{red}{reservoir-block construction}, representative error model, robustness scan, and synthetic tomography are
given in the SM. %The analysis is theoretical, and experimental validation remains to be done.

\textit{Summary---}We define the entanglement Mpemba effect by an ordering reversal between two
entanglement trajectories under the same open-system dynamics.  The
entanglement-visible comparison spectrum isolates the modes that determine the
late-time order and gives a sufficient crossing criterion even when the
individual trajectories are nonmonotonic. Furthermore, deterministic LOCC
reachability supplies an all-monotone ordering certificate for general mixed and
multipartite states, %and product-source Bell and GHZ variants realize itsreversal under fixed Lindbladians.}
In the original solvable Bell- and GHZ-state pumps, the initial state selects
a kinetic sector and thereby controls both the reversal and the hitting time.
The fixed-cycle construction changes only the initial state.

The spectral criterion concerns one chosen entanglement monotone in a common
asymptotic sector. The LOCC certificate is stronger, assuring a universal entanglement order reversal. Near the attractor,
visible rates and grouped coefficients can compare initial states without
reconstructing their full trajectories.  In our examples the common nonlocal
reservoir supplies entanglement, and unequal access to its kinetic sectors
produces the inversion.  The mechanism allows nonmonotonicity, kinks, and
multiple crossings.  Device-calibrated tests, degenerate stationary
manifolds, and closing many-body Liouvillian gaps remain open.

\paragraph*{\textit{Note added.}}
On the day we submitted this manuscript, we became aware of the independent
work of Benjadi \textit{et al.}~\cite{Benjadi2026}, which also studies
Mpemba-enhanced entanglement generation. %Our work additionally provides a mixed-state LOCC-order realization and gives an explicit genuine-multipartite realization.

% Add acknowledgments and funding information before submission.

% Include the reference used only in the Supplemental Material.
\nocite{Plenio2005}
\bibliography{ref}

@article{mpemba1969cool,
  title = {{Cool?}},
  author = {Mpemba, Erasto B. and Osborne, Denis G.},
  journal = {Phys. Educ.},
  volume = {4},
  number = {3},
  pages = {172--175},
  year = {1969},
  doi = {10.1088/0031-9120/4/3/312},
  url = {https://iopscience.iop.org/article/10.1088/0031-9120/4/3/312},
  publisher = {IOP Publishing}
}

@article{Torrente19,
  title = {{Large Mpemba-like effect in a gas of inelastic rough hard spheres}},
  author = {Torrente, Aurora and L{\'o}pez-Casta{\~n}o, Miguel A. and Lasanta, Antonio and Reyes, Francisco Vega and Prados, Antonio and Santos, Andr{\'e}s},
  journal = {Phys. Rev. E},
  volume = {99},
  number = {6},
  pages = {060901},
  year = {2019},
  month = {Jun},
  publisher = {American Physical Society},
  doi = {10.1103/PhysRevE.99.060901}
}

@article{Biswas20,
  title = {{Mpemba effect in driven granular Maxwell gases}},
  author = {Biswas, Apurba and Prasad, V. V. and Raz, O. and Rajesh, R.},
  journal = {Phys. Rev. E},
  volume = {102},
  number = {1},
  pages = {012906},
  year = {2020},
  month = {Jul},
  publisher = {American Physical Society},
  doi = {10.1103/PhysRevE.102.012906}
}

@article{Santos20,
  title = {{Mpemba effect in molecular gases under nonlinear drag}},
  author = {Santos, Andr{\'e}s and Prados, Antonio},
  journal = {Phys. Fluids},
  volume = {32},
  number = {7},
  pages = {072010},
  year = {2020},
  month = {Jul},
  publisher = {AIP Publishing},
  doi = {10.1063/5.0016243}
}

@article{kumar2020exponentially,
  title = {{Exponentially faster cooling in a colloidal system}},
  author = {Kumar, Avinash and Bechhoefer, John},
  journal = {Nature},
  volume = {584},
  number = {7819},
  pages = {64--68},
  year = {2020},
  url = {https://www.nature.com/articles/s41586-020-2560-x},
  publisher = {Nature Portfolio}
}

@article{Ivander23,
  title = {{Hyperacceleration of quantum thermalization dynamics by bypassing long-lived coherences: An analytical treatment}},
  author = {Ivander, Felix and Anto-Sztrikacs, Nicholas and Segal, Dvira},
  journal = {Phys. Rev. E},
  volume = {108},
  number = {1},
  pages = {014130},
  year = {2023},
  month = {Jul},
  publisher = {American Physical Society},
  doi = {10.1103/PhysRevE.108.014130}
}

@article{Carollo21,
  title = {{Exponentially Accelerated Approach to Stationarity in Markovian Open Quantum Systems through the Mpemba Effect}},
  author = {Carollo, Federico and Lasanta, Antonio and Lesanovsky, Igor},
  journal = {Phys. Rev. Lett.},
  volume = {127},
  number = {6},
  pages = {060401},
  year = {2021},
  month = {Aug},
  publisher = {American Physical Society},
  doi = {10.1103/PhysRevLett.127.060401}
}

@article{Biswas24,
  title = {{Mpemba effect on nonequilibrium active Markov chains}},
  author = {Biswas, Apurba and Pal, Arnab},
  journal = {Phys. Rev. E},
  volume = {111},
  number = {5},
  pages = {054136},
  year = {2025},
  month = {May},
  publisher = {American Physical Society},
  doi = {10.1103/PhysRevE.111.054136}
}

@article{lu2017nonequilibrium,
  title = {{Nonequilibrium thermodynamics of the Markovian Mpemba effect and its inverse}},
  author = {Lu, Zhiyue and Raz, Oren},
  journal = {Proc. Natl. Acad. Sci. U.S.A.},
  volume = {114},
  number = {20},
  pages = {5083--5088},
  year = {2017},
  doi = {10.1073/pnas.1701264114},
  url = {https://www.pnas.org/doi/10.1073/pnas.1701264114},
  publisher = {National Academy of Sciences}
}

@article{Gonzalez2021,
  title = {{Mpemba-like effect in driven binary mixtures}},
  author = {G{\'o}mez Gonz{\'a}lez, Rub{\'e}n and Khalil, Nagi and Garz{\'o}, Vicente},
  journal = {Phys. Fluids},
  volume = {33},
  number = {5},
  pages = {053301},
  year = {2021},
  month = {May},
  publisher = {AIP Publishing},
  doi = {10.1063/5.0050530}
}

@article{liu2026mpemba,
  title={The {M}pemba effect likes to hit a wall},
  author={Liu, Yue and Van Vu, Tan and Ch{\'e}trite, Rapha{\"e}l and van Wijland, Fr{\'e}d{\'e}ric and Hayakawa, Hisao},
  journal={arXiv preprint arXiv:2604.01543},
  year={2026}
}

@article{pre-us,
  title = {{Predicting the conditions for observing the Mpemba effect}},
  author = {Liu, Yue and Van Vu, Tan and Ch{\'e}trite, Rapha{\"e}l and van Wijland, Fr{\'e}d{\'e}ric and Hayakawa, Hisao},
  journal = {arXiv preprint arXiv:2606.03445},
  url = {https://arxiv.org/abs/2606.03445},
  year = {2026}
}

@article{klich2019mpemba,
  title = {{Mpemba index and anomalous relaxation}},
  author = {Klich, Israel and Raz, Oren and Hirschberg, Ori and Vucelja, Marija},
  journal = {Phys. Rev. X},
  volume = {9},
  number = {2},
  pages = {021060},
  year = {2019},
  doi = {10.1103/PhysRevX.9.021060},
  url = {https://journals.aps.org/prx/abstract/10.1103/PhysRevX.9.021060},
  publisher = {American Physical Society},
}

@article{walker2021anomalous,
  title = {{Anomalous thermal relaxation of Langevin particles in a piecewise-constant potential}},
  author = {Walker, Matthew R. and Vucelja, Marija},
  journal = {J. Stat. Mech.},
  volume = {2021},
  number = {11},
  pages = {113105},
  year = {2021},
  publisher = {IOP Publishing},
  url = {https://iopscience.iop.org/article/10.1088/1742-5468/ac2edc},
}

@article{chetrite2021metastable,
  title = {{The metastable Mpemba effect corresponds to a non-monotonic temperature dependence of extractable work}},
  author = {Ch{\'e}trite, Rapha{\"e}l and Kumar, Avinash and Bechhoefer, John},
  journal = {Front. Phys.},
  volume = {9},
  pages = {654271},
  year = {2021},
  url = {https://www.frontiersin.org/journals/physics/articles/10.3389/fphy.2021.654271/full},
  publisher = {Frontiers Media SA}
}

@article{Takada21a,
  title = {{Mpemba effect in inertial suspensions}},
  author = {Takada, Satoshi and Hayakawa, Hisao and Santos, Andr{\'e}s},
  journal = {Phys. Rev. E},
  volume = {103},
  number = {3},
  pages = {032901},
  year = {2021},
  month = {Mar},
  publisher = {American Physical Society},
  doi = {10.1103/PhysRevE.103.032901}
}

@article{Santos17,
  title = {{When the Hotter Cools More Quickly: Mpemba Effect in Granular Fluids}},
  author = {Lasanta, Antonio and Vega Reyes, Francisco and Prados, Antonio and Santos, Andr{\'e}s},
  journal = {Phys. Rev. Lett.},
  volume = {119},
  number = {14},
  pages = {148001},
  year = {2017},
  month = {Oct},
  publisher = {American Physical Society},
  doi = {10.1103/PhysRevLett.119.148001}
}

@article{Kumar22,
  title = {{Anomalous heating in a colloidal system}},
  author = {Kumar, Avinash and Ch{\'e}trite, Rapha{\"e}l and Bechhoefer, John},
  journal = {Proc. Natl. Acad. Sci. U.S.A.},
  volume = {119},
  number = {5},
  pages = {e2118484119},
  year = {2022},
  doi = {10.1073/pnas.2118484119}
}

@article{Moroder2024,
  title = {{Thermodynamics of the Quantum Mpemba Effect}},
  author = {Moroder, Mattia and Culhane, Ois{\'i}n and Zawadzki, Krissia and Goold, John},
  journal = {Phys. Rev. Lett.},
  volume = {133},
  number = {14},
  pages = {140404},
  year = {2024},
  month = {Oct},
  publisher = {American Physical Society},
  doi = {10.1103/PhysRevLett.133.140404}
}

@article{Joshi2024,
  title = {{Observing the Quantum Mpemba Effect in Quantum Simulations}},
  author = {Joshi, Lata Kh. and Franke, Johannes and Rath, Aniket and Ares, Filiberto and Murciano, Sara and Kranzl, Florian and Blatt, Rainer and Zoller, Peter and Vermersch, Beno{\^\i}t and Calabrese, Pasquale and Roos, Christian F. and Joshi, Manoj K.},
  journal = {Phys. Rev. Lett.},
  volume = {133},
  number = {1},
  pages = {010402},
  year = {2024},
  month = {Jul},
  publisher = {American Physical Society},
  doi = {10.1103/PhysRevLett.133.010402}
}

@article{Chatterjee2024,
  title = {{Multiple quantum Mpemba effect: Exceptional points and oscillations}},
  author = {Chatterjee, Amit Kumar and Takada, Satoshi and Hayakawa, Hisao},
  journal = {Phys. Rev. A},
  volume = {110},
  number = {2},
  pages = {022213},
  year = {2024},
  month = {Aug},
  publisher = {American Physical Society},
  doi = {10.1103/PhysRevA.110.022213}
}

@article{Chatterjee_2023,
  title = {{Quantum Mpemba Effect in a Quantum Dot with Reservoirs}},
  author = {Chatterjee, Amit Kumar and Takada, Satoshi and Hayakawa, Hisao},
  journal = {Phys. Rev. Lett.},
  volume = {131},
  number = {8},
  pages = {080402},
  year = {2023},
  month = {Aug},
  publisher = {American Physical Society},
  doi = {10.1103/PhysRevLett.131.080402}
}

@article{Shapira2024,
  title = {{Inverse Mpemba Effect Demonstrated on a Single Trapped Ion Qubit}},
  author = {Shapira, Shahaf Aharony and Shapira, Yotam and Markov, Jovan and Teza, Gianluca and Akerman, Nitzan and Raz, Oren and Ozeri, Roee},
  journal = {Phys. Rev. Lett.},
  volume = {133},
  number = {1},
  pages = {010403},
  year = {2024},
  month = {Jul},
  publisher = {American Physical Society},
  doi = {10.1103/PhysRevLett.133.010403}
}

@article{Wang24,
  title = {{Mpemba effects in nonequilibrium open quantum systems}},
  author = {Wang, Xuanhua and Wang, Jin},
  journal = {Phys. Rev. Res.},
  volume = {6},
  number = {3},
  pages = {033330},
  year = {2024},
  month = {Sep},
  publisher = {American Physical Society},
  doi = {10.1103/PhysRevResearch.6.033330}
}

@article{Nava24,
  title = {{Mpemba Effects in Open Nonequilibrium Quantum Systems}},
  author = {Nava, Andrea and Egger, Reinhold},
  journal = {Phys. Rev. Lett.},
  volume = {133},
  number = {13},
  pages = {136302},
  year = {2024},
  month = {Sep},
  publisher = {American Physical Society},
  doi = {10.1103/PhysRevLett.133.136302}
}

@article{Tan2025,
  title = {{Thermomajorization Mpemba Effect}},
  author = {Van Vu, Tan and Hayakawa, Hisao},
  journal = {Phys. Rev. Lett.},
  volume = {134},
  number = {10},
  pages = {107101},
  year = {2025},
  month = {Mar},
  publisher = {American Physical Society},
  doi = {10.1103/PhysRevLett.134.107101}
}

@article{Strachan2025,
  title = {{Non-Markovian Quantum Mpemba Effect}},
  author = {Strachan, David J. and Purkayastha, Archak and Clark, Stephen R.},
  journal = {Phys. Rev. Lett.},
  volume = {134},
  number = {22},
  pages = {220403},
  year = {2025},
  month = {Jun},
  publisher = {American Physical Society},
  doi = {10.1103/PhysRevLett.134.220403}
}

@article{Turkeshi2025,
  title = {{Quantum Mpemba Effect in Random Circuits}},
  author = {Turkeshi, Xhek and Calabrese, Pasquale and De Luca, Andrea},
  journal = {Phys. Rev. Lett.},
  volume = {135},
  number = {4},
  pages = {040403},
  year = {2025},
  month = {Jul},
  publisher = {American Physical Society},
  doi = {10.1103/5d6p-8d1b}
}

@article{Zhang2025,
  title = {{Observation of quantum strong Mpemba effect}},
  author = {Zhang, Jie and Xia, Gang and Wu, Chun-Wang and Chen, Ting and Zhang, Qian and Xie, Yi and Su, Wen-Bo and Wu, Wei and Qiu, Cheng-Wei and Chen, Ping-Xing and Li, Weibin and Jing, Hui and Zhou, Yan-Li},
  journal = {Nat. Commun.},
  volume = {16},
  number = {1},
  pages = {301},
  year = {2025},
  month = {Jan},
  publisher = {Nature Portfolio},
  doi = {10.1038/s41467-024-54303-0}
}

@article{Bao2025,
  title = {{Accelerating Quantum Relaxation via Temporary Reset: A Mpemba-Inspired Approach}},
  author = {Bao, Ruicheng and Hou, Zhonghuai},
  journal = {Phys. Rev. Lett.},
  volume = {135},
  number = {15},
  pages = {150403},
  year = {2025},
  month = {Oct},
  publisher = {American Physical Society},
  doi = {10.1103/g94p-7421}
}

@article{Bao2026,
  title = {{Initial-State Typicality in Quantum Relaxation}},
  author = {Bao, Ruicheng},
  journal = {Phys. Rev. Lett.},
  volume = {136},
  number = {7},
  pages = {070402},
  year = {2026},
  month = {Feb},
  publisher = {American Physical Society},
  doi = {10.1103/wgr5-lb6b}
}

@article{TEZA2026,
  title = {{Speedups in nonequilibrium thermal relaxation: Mpemba and related effects}},
  author = {Teza, Gianluca and Bechhoefer, John and Lasanta, Antonio and Raz, Oren and Vucelja, Marija},
  journal = {Phys. Rep.},
  volume = {1164},
  pages = {1--97},
  year = {2026},
  publisher = {Elsevier},
  doi = {10.1016/j.physrep.2025.002984}
}

@article{Degunther2022,
doi = {10.1209/0295-5075/ac8573},
url = {https://doi.org/10.1209/0295-5075/ac8573},
year = {2022},
month = {aug},
publisher = {EDP Sciences, IOP Publishing and Societ{\`a} Italiana di Fisica},
volume = {139},
number = {4},
pages = {41002},
author = {Deg\"{u}nther, Julius and Seifert, Udo},
title = {Anomalous relaxation from a non-equilibrium steady state: An isothermal analog of the Mpemba effect},
journal = {Europhys. Lett.},
}

@Article{Yu2025,
author={Yu, Hui
and Liu, Shuo
and Zhang, Shi-Xin},
title={Quantum Mpemba effects from symmetry perspectives},
journal={AAPPS Bulletin},
year={2025},
month={Jul},
day={02},
volume={35},
number={1},
pages={17},
issn={2309-4710},
doi={10.1007/s43673-025-00157-7},
url={https://doi.org/10.1007/s43673-025-00157-7}
}

@article{AresReview2025,
  author = {Ares, Filiberto and Calabrese, Pasquale and Murciano, Sara},
  title = {The quantum Mpemba effects},
  journal = {Nat. Rev. Phys.},
  volume = {7},
  pages = {451--460},
  year = {2025},
  doi = {10.1038/s42254-025-00838-0}
}

@article{Wootters1998,
  author = {Wootters, William K.},
  title = {Entanglement of Formation of an Arbitrary State of Two Qubits},
  journal = {Phys. Rev. Lett.},
  volume = {80},
  pages = {2245--2248},
  year = {1998},
  doi = {10.1103/PhysRevLett.80.2245}
}

@article{VidalWerner2002,
  author = {Vidal, Guifr{\'e} and Werner, Reinhard F.},
  title = {Computable measure of entanglement},
  journal = {Phys. Rev. A},
  volume = {65},
  pages = {032314},
  year = {2002},
  doi = {10.1103/PhysRevA.65.032314}
}

@article{Plenio2005,
  author = {Plenio, Martin B.},
  title = {Logarithmic Negativity: A Full Entanglement Monotone That Is Not Convex},
  journal = {Phys. Rev. Lett.},
  volume = {95},
  pages = {090503},
  year = {2005},
  doi = {10.1103/PhysRevLett.95.090503}
}

@article{Horodecki2009,
  author = {Horodecki, Ryszard and Horodecki, Pawe{\l} and Horodecki, Micha{\l} and Horodecki, Karol},
  title = {Quantum entanglement},
  journal = {Rev. Mod. Phys.},
  volume = {81},
  pages = {865--942},
  year = {2009},
  doi = {10.1103/RevModPhys.81.865}
}

@article{Nielsen1999,
  author = {Nielsen, Michael A.},
  title = {Conditions for a Class of Entanglement Transformations},
  journal = {Phys. Rev. Lett.},
  volume = {83},
  pages = {436--439},
  year = {1999},
  doi = {10.1103/PhysRevLett.83.436}
}

@article{Vidal2000,
  author = {Vidal, Guifr{\'e}},
  title = {Entanglement monotones},
  journal = {J. Mod. Opt.},
  volume = {47},
  pages = {355--376},
  year = {2000},
  doi = {10.1080/09500340008244048}
}

@article{Rafsanjani2012,
  author = {Hashemi Rafsanjani, S. M. and Huber, M. and Broadbent, C. J. and Eberly, J. H.},
  title = {Genuine multipartite concurrence of {$N$}-qubit {$X$} matrices},
  journal = {Phys. Rev. A},
  volume = {86},
  pages = {062303},
  year = {2012},
  doi = {10.1103/PhysRevA.86.062303}
}

@article{Lindblad1976,
  author = {Lindblad, G{\"o}ran},
  title = {On the generators of quantum dynamical semigroups},
  journal = {Commun. Math. Phys.},
  volume = {48},
  pages = {119--130},
  year = {1976},
  doi = {10.1007/BF01608499}
}

@article{Gorini1976,
  author = {Gorini, Vittorio and Kossakowski, Andrzej and Sudarshan, E. C. G.},
  title = {Completely positive dynamical semigroups of {$N$}-level systems},
  journal = {J. Math. Phys.},
  volume = {17},
  pages = {821--825},
  year = {1976},
  doi = {10.1063/1.522979}
}

@article{Barreiro2011,
  author = {Barreiro, Julio T. and M{\"u}ller, Markus and Schindler, Philipp and Nigg, Daniel and Monz, Thomas and Chwalla, Michael and Hennrich, Markus and Roos, Christian F. and Zoller, Peter and Blatt, Rainer},
  title = {An open-system quantum simulator with trapped ions},
  journal = {Nature},
  volume = {470},
  pages = {486--491},
  year = {2011},
  doi = {10.1038/nature09801}
}

@article{Mueller2011,
  author = {M{\"u}ller, Markus and Hammerer, Klemens and Zhou, Y. L. and Roos, Christian F. and Zoller, Peter},
  title = {Simulating open quantum systems: From many-body interactions to stabilizer pumping},
  journal = {New J. Phys.},
  volume = {13},
  pages = {085007},
  year = {2011},
  doi = {10.1088/1367-2630/13/8/085007}
}

@article{Schindler2013,
  author = {Schindler, Philipp and M{\"u}ller, Markus and Nigg, Daniel and Barreiro, Julio T. and Martinez, Esteban A. and Hennrich, Markus and Monz, Thomas and Diehl, Sebastian and Zoller, Peter and Blatt, Rainer},
  title = {Quantum simulation of dynamical maps with trapped ions},
  journal = {Nat. Phys.},
  volume = {9},
  pages = {361--367},
  year = {2013},
  doi = {10.1038/nphys2630}
}

@article{Lin2013,
  author = {Lin, Y. and Gaebler, J. P. and Reiter, F. and Tan, T. R. and Bowler, R. and S{\o}rensen, A. S. and Leibfried, D. and Wineland, D. J.},
  title = {Dissipative production of a maximally entangled steady state of two quantum bits},
  journal = {Nature},
  volume = {504},
  pages = {415--418},
  year = {2013},
  doi = {10.1038/nature12801}
}

@article{MolmerSorensen1999,
  author = {M{\o}lmer, Klaus and S{\o}rensen, Anders},
  title = {Multiparticle Entanglement of Hot Trapped Ions},
  journal = {Phys. Rev. Lett.},
  volume = {82},
  pages = {1835--1838},
  year = {1999},
  doi = {10.1103/PhysRevLett.82.1835}
}

@article{Bennett1993,
  author = {Bennett, Charles H. and Brassard, Gilles and Cr{\'e}peau, Claude and Jozsa, Richard and Peres, Asher and Wootters, William K.},
  title = {Teleporting an Unknown Quantum State via Dual Classical and {Einstein--Podolsky--Rosen} Channels},
  journal = {Phys. Rev. Lett.},
  volume = {70},
  pages = {1895--1899},
  year = {1993},
  doi = {10.1103/PhysRevLett.70.1895}
}

@article{Ekert1991,
  author = {Ekert, Artur K.},
  title = {Quantum Cryptography Based on {Bell's} Theorem},
  journal = {Phys. Rev. Lett.},
  volume = {67},
  pages = {661--663},
  year = {1991},
  doi = {10.1103/PhysRevLett.67.661}
}

@article{Raussendorf2001,
  author = {Raussendorf, Robert and Briegel, Hans J.},
  title = {A One-Way Quantum Computer},
  journal = {Phys. Rev. Lett.},
  volume = {86},
  pages = {5188--5191},
  year = {2001},
  doi = {10.1103/PhysRevLett.86.5188}
}

@article{Kimble2008,
  author = {Kimble, H. J.},
  title = {The Quantum Internet},
  journal = {Nature},
  volume = {453},
  pages = {1023--1030},
  year = {2008},
  doi = {10.1038/nature07127}
}

@article{Giovannetti2011,
  author = {Giovannetti, Vittorio and Lloyd, Seth and Maccone, Lorenzo},
  title = {Advances in Quantum Metrology},
  journal = {Nat. Photonics},
  volume = {5},
  pages = {222--229},
  year = {2011},
  doi = {10.1038/nphoton.2011.35}
}

@article{Verstraete2009,
  author = {Verstraete, Frank and Wolf, Michael M. and Cirac, J. Ignacio},
  title = {Quantum Computation and Quantum-State Engineering Driven by Dissipation},
  journal = {Nat. Phys.},
  volume = {5},
  pages = {633--636},
  year = {2009},
  doi = {10.1038/nphys1342}
}

@article{Pocklington2024,
  author = {Pocklington, Andrew and Clerk, Aashish A.},
  title = {Universal Time-Entanglement Trade-Off in Open Quantum Systems},
  journal = {PRX Quantum},
  volume = {5},
  pages = {040305},
  year = {2024},
  doi = {10.1103/PRXQuantum.5.040305}
}

@article{Pocklington2025,
  author = {Pocklington, Andrew and Clerk, Aashish A.},
  title = {Accelerating Dissipative State Preparation with Adaptive Open Quantum Dynamics},
  journal = {Phys. Rev. Lett.},
  volume = {134},
  pages = {050603},
  year = {2025},
  doi = {10.1103/PhysRevLett.134.050603}
}

@article{Kraus2008,
  author = {Kraus, B. and B{\"u}chler, H. P. and Diehl, S. and Kantian, A. and Micheli, A. and Zoller, P.},
  title = {Preparation of Entangled States by Quantum Markov Processes},
  journal = {Phys. Rev. A},
  volume = {78},
  number = {4},
  pages = {042307},
  year = {2008},
  month = {Oct},
  publisher = {American Physical Society},
  doi = {10.1103/PhysRevA.78.042307},
  url = {https://doi.org/10.1103/PhysRevA.78.042307}
}

@article{Pandey2024,
  author = {Pandey, Vivek and Bhowmick, Swapnil and Mohan, Brij and Sohail and Sen, Ujjwal},
  title = {Fundamental Speed Limits on Entanglement Dynamics of Bipartite Quantum Systems},
  journal = {Phys. Rev. A},
  volume = {110},
  number = {5},
  pages = {052420},
  year = {2024},
  month = {Nov},
  publisher = {American Physical Society},
  doi = {10.1103/PhysRevA.110.052420},
  url = {https://doi.org/10.1103/PhysRevA.110.052420}
}

@article{Sorelli2019,
  author = {Sorelli, Giacomo and Gessner, Manuel and Smerzi, Augusto and Pezz{\`e}, Luca},
  title = {Fast and Optimal Generation of Entanglement in Bosonic Josephson Junctions},
  journal = {Phys. Rev. A},
  volume = {99},
  number = {2},
  pages = {022329},
  year = {2019},
  month = {Feb},
  publisher = {American Physical Society},
  doi = {10.1103/PhysRevA.99.022329},
  url = {https://doi.org/10.1103/PhysRevA.99.022329}
}

@article{LiuNie2023,
  author = {Liu, Junjie and Nie, Hanlin},
  title = {Initial-State-Dependent Quantum Speed Limit for Dissipative State Preparation: Framework and Optimization},
  journal = {Phys. Rev. A},
  volume = {107},
  number = {5},
  pages = {052608},
  year = {2023},
  month = {May},
  publisher = {American Physical Society},
  doi = {10.1103/PhysRevA.107.052608},
  url = {https://doi.org/10.1103/PhysRevA.107.052608}
}

@article{Benjadi2026,
  title = {{Exponential Speedup of Entanglement Generation by Quantum Mpemba Effects}},
  author = {Benjadi, Sara M. and Egger, Reinhold and Gornyi, Igor and Nava, Andrea},
  journal = {arXiv preprint arXiv:2608.05935},
  year = {2026},
  month = {Aug},
  url = {https://arxiv.org/abs/2608.05935}
}

@misc{SupplementalMaterial,
  note = {See Supplemental Material at [URL will be inserted by publisher] for detailed derivations}
}

\end{document}